\documentclass[12pt,preprint]{aastex}
\shorttitle{Photometric variability of the peculiar magnetic white dwarf WD\,1953-011}
\shortauthors{Valyavin et al.}
\begin{document}

\title{A study of the photometric variability of the peculiar magnetic white dwarf WD\,1953-011
}

\author{G.~Valyavin\altaffilmark{1}, K.~Antonyuk\altaffilmark{2},
S.Plachinda\altaffilmark{2},
D.M.~Clark\altaffilmark{1},
G.A.~Wade\altaffilmark{3},
L.~Fox Machado\altaffilmark{1},
M.~Alvarez\altaffilmark{1}
J.M.~Lopez\altaffilmark{1}
D.~Hiriart\altaffilmark{1}
Inwoo Han\altaffilmark{4},
Young-Beom Jeon\altaffilmark{4},
S.~Bagnulo\altaffilmark{5},
S.V.~Zharikov\altaffilmark{1},
C.~Zurita\altaffilmark{6},
R.~Mujica\altaffilmark{7}
D.~Shulyak\altaffilmark{8}
T.~Burlakova\altaffilmark{9}
}

\altaffiltext{1}{
Observatorio Astron\'{o}mico Nacional SPM,
Instituto de Astronom\'{i}a, Universidad Nacional Aut\'{o}noma de M\'{e}xico,
Ensenada, BC, M\'{e}xico}
\altaffiltext{2}{Crimean Astrophysical Observatory, Nauchny, Crimea, Ukraine}
\altaffiltext{3}{Physics Department, Royal
Military College of Canada, Kingston, Ontario, Canada}
\altaffiltext{4}{Korea Astronomy and Space Science Institute, 61-1,
Whaam-Dong, Youseong-Gu, Daejeon, Republic of Korea 305-348 }
\altaffiltext{5}{Armagh Observatory, Northern Ireland}
\altaffiltext{6}{
Instituto de Astrof\'{i}sica de Canarias, 38200, La Laguna, Tenerife, Spain
and Departamento de Astrof\'{i}sica, Universidad de La Laguna, La Laguna,
Tenerife, Spain}
\altaffiltext{7}{Instituto Nacional de Astrof\'{i}sica, \'{O}ptica
y Electron\'{i}ca, Apartado Postal 51 y 216, 72000, Tonantzintla, Pueblo,
M\'{e}xico}
\altaffiltext{8}{Astrophysics, Georg-August-University, Friedrich-Hund-Platz
1, D-37077 G\"ottingen, Germany}
\altaffiltext{9}{Special Astrophysical Observatory, Zelenchukskaya,
Russia}

\begin{abstract}
We present and interpret simultaneous new photometric and spectroscopic
observations of the peculiar magnetic white dwarf WD\,1953-011.
The flux in the V-band filter and intensity of the Balmer
spectral lines demonstrate variability with the rotation period of about 1.45~days.
According to previous studies, this variability can be explained by the
presence of a dark spot having a magnetic nature, analogous to a sunspot.
Motivated by this idea, we examine possible physical relationships between
the suggested dark spot and the strong-field magnetic structure (magnetic ``spot'', or ``tube'') recently identified on the
surface of this star. Comparing the rotationally-modulated flux
with the variable spectral observables related to the magnetic ``spot'' we
establish their correlation, and therefore their physical relationship.
Modeling the variable photometric flux assuming that it is associated with temperature variations in the stellar photosphere,
we argue that the strong-field area and dark, low-temperature spot are
comparable in size and located at the same latitudes, essentially
overlapping each other with a possible slight longitudinal shift. 
In this paper we also present a new, improved value of the star's rotational period and constrain the characteristics of the thermal inhomogeneity over the degenerate's surface.
\end{abstract}

\section{Introduction}
\label{Intro}

The magnetic white dwarf WD1953-011 exhibits an unusual magnetic field geometry,
consisting of two qualitatively different morphological components - a weak,
global dipole field, and a strong, localized magnetic spot
\citep{MFMW00,Wad03,VWB08}. The mean field modulus of the spot,
$\approx 500$\,kG \citep{MFMW00,VWB08}, and its variable with rotation
period longitudinal component suggest dominating vertical orientation of the
magnetic field lines in most of the spot's area. This makes it possible to
interpret this feature as a huge magnetic flux tube covering about 20\%
of the star's visible hemisphere \citep{VWB08}. If this suggestion is correct,
we may expect some other observational effects related to the magnetic spot.
For example, according to the basic properties of magnetic fields in stars
(for instance, \citet{PA79}), and by analogy to sunspots,
such localized fields may have a significant impact on the
pressure-temperature balance in the photospheres of stars. This may produce
a temperature difference between the strong-field area and other parts of
the star's surface. As a result we may expect rotationally-modulated
photometric variability of WD\,1953-011. For the same reasons such fields
might be unstable and exhibit secular drift over the stellar surface.

In this context, we point out that significant photometric variability of
WD\,1953-011 has been established \citep{Wad03,BM05}. The results of
high-resolution spectroscopy \citep{MFMW00} make it difficult
to explain the observed flux variations as a consequence of binarity.
Non-thermal effects such as the generation of rotationally-modulated synchrotron
radiation also provide no arguments for the photometric variability,
due to the comparatively weak magnetic field of the star and absence of
relativistic free electrons of sufficient concentration.
We therefore suppose that the only reasonable explanation for these
variations is surface thermal inhomogeneities,
which might form a group of dark spots, or one single spot \citep{BM05}.
As well, some arguments for the presence of secular drifts of the
strong-field and dark area have also been discussed
\citep{Wad03,VWB08}.

Indirectly, these considerations suggest a physical relationship between the dark
and magnetic spots, and their possible secular migration. However, our
previous spectroscopy and available photometry were obtained at different
epochs, making it impossible to study this relationship in detail.
Examination of these problems is the goal of the current study,
based on additional {\it simultaneous} photometric
and spectroscopic observations of WD\,1953-011.

\section{Observations}
\label{Obs}

The main portion of the simultaneous spectral and photometric observations of WD~1953$-$011
were carried out at the Crimean Astrophysical Observatory (CRAO, Ukraine)
during several observing nights between July and August 2007. The
low resolution long-slit spectrograph of the 2.6\,m telescope ZTSh and the UBVR CCD
photometer at the 1.25\,m telescope AZT-11 were used in observations at CRAO.
The spectrograph used is a standard instrument intended for low-resolution
spectroscopy, and is described by \citet{Dor08}. The spectrograph was used
to obtain spectra of the H$\beta$ region with the grating 1200\,gv/mm
providing a wavelength coverage of 4330\,\AA -- 5330\,\AA. With a slit
width of 3\arcsec , the spectral resolving power was about 3.8$\AA$ (R1300).

Photometric observations were carried out with the Ritchey-Chretien AZT-11 telescope
(D = 1.25m; F = 16m). 
The KAF-1001E CCD-camera (Finger Lakes Instrumentation company), with
pixel size - 24 x 24 $\mu$m, frame size - 1k$\times$1k pixels,
and standard Johnson V-band filter, were used in these observations.

These observations were also supported by a few observing runs during
May-June, 2009, with the 2.1-m telescope at the
Observatorio Astron\'{o}mico Nacional (Mexico: OAN\,SPM). With this instrument
we obtained an additional series of spectra of WD\,1953-011 using the
moderate dispersion echelle spectrograph REOSC with a resolving power of about 17000.
The instrument is described in detail by \citet[see also the website of
the OAN\,SPM: www.astrossp.unam.mx/indexspm.html]{LC94}.
A series of additional photometric observations were also obtained with the
standard CCD photometers at the 0.84- and 1.5-m telescopes at SPM (see section
``instruments'' www.astrossp.unam.mx/indexspm.html ).
In this study we used the REOSC in the low resolution regime (binnig spectral
material to a resolution of about R=3000).

Standard observational techniques and data reduction were used.
In order to analyse the observations
of WD1953-011 in combination with observations of previous authors
\citep{Wad03,BM05,VWB08}, all observational and data processing steps
were made similar to those described by those authors.
As summary of the new spectral and photometric observations is presented in
Table~\ref{tbl1} and Table~\ref{tbl2}, where we also present results of
measurements as explained below.

\section{Measurements}

In this section we describe the measurements necessary for analysis of our data.
Due to the specific field morphology of WD\,1953-011 we begin with some
qualitative explanations on the observables we used.

According to previous studies, the magnetic field of
WD\,1953-011 consists of two distinct components -- a low-field global dipole +
quadrupole component \citep{VWB08}  superimposed with the localized
strong-field area \citep{MFMW00}. The low-field ($\sim 100$\,kG) component
of the degenerate was discovered spectroscopically by \citet{KDWA98}
via analysis of the Zeeman pattern in the H$\alpha$ core.
The strong-field ``spot'' was detected by \citet{MFMW00} via observations of
additional strong-field Zeeman features in the H$\alpha$ wings.
Both patterns are variable due to the star's rotation, and can be measured independently in high-resolution
spectroscopy. Together with the variable flux, these are the most important
observables.

Due to the low spectral resolution and S/N of our current spectral observations,  (in contrast to that which
we used in our previous studies of this degenerate with the 8-m telescope
at VLT) here we are unable to provide a robust analysis of the
star's global magnetic field revealed in cores of the Balmer lines.
However, the strong-field $\approx 0.5\,MG$ magnetic spot can be effectively studied
in terms of residual intensities and equivalent widths of the Balmer lines.

The Balmer line profiles in the spectrum of WD\,1953-011 vary due to
the strong-field magnetic spot. The central intensity $r_c$ of the line cores,
and the equivalent widths $EW$ of the lines, correlate strongly with the
intensity of the strong-field Zeeman features which are found in the wings of
the Balmer lines (for example, see
\citet{VWB08} and Fig.~3 therein: the two satellite features are at
$\pm$10~\AA \, around the H$\alpha$ core). This correlation (the higher the
intensity of the features, the weaker $r_c$ and $EW$,
see Fig.~3 in \citet{VWB08}) is due to the
fact that the spot, which appears periodically on the visible disc due to rotation,
significantly redistributes the H$\alpha$ flux according to its projected area.
Taking these arguments into account, we conclude that the minimum of $r_c$
and $EW$ of the Balmer lines correspond to those moments (or close
to those moments), when the strong-field magnetic area is projected towards
the observer. The minimum projection of the strong-field area corresponds to
the deepest central intensities and largest equivalent widths of the Balmer
lines. Due to this, in this paper we use $r_c$ and $EW$ of the star's
Balmer spectral lines as the main observables to examine the rotational
modulation of the strong-field area, and to
compare with the flux variability.

At the same time it is important to note that both the observables
$r_c$ and $EW$ are also sensitive to thermal effects at the star's surface
and this may introduce some uncertainties in our analysis. However, empirical
examination of our previous, high signal-to-noise spectropolarimetric
observations of this star with the VLT made it possible to accept the above
assumption as a good first-guess approximation. Presented by
\citet{VWB08} analysis of rotationally
modulated circular polarization in the H$\alpha$ wings also supports this
assumption, revealing a direct empirical relationship between the amplitudes of
circular polarization attributed to the strong-field area and $r_c$ or $EW$
(the higher the circular polarization the lower $r_c$ and $EW$).

\subsection{Spectroscopic measurements}

Taking the above arguments into account we measured the $EW$ and $r_c$ of
the H$\beta$ spectral line in observations with ZTSH and H$\alpha$
in all other cases. In measurements of the equivalent widths
we used all parts of the profiles including the satellite sigma components
within the window of $\pm 50 $\,\AA\,
from the Balmer line cores. In order to minimize selection
effects due to the use of different telescopes, spectrographs and different
spectral lines, all the measurements within individual groups of observations
with a given instrument and spectral line were normalized by their
mean values, averaged over a full rotational cycle of the star. For example,
normalized $EW$ values of the H$\alpha$ profile obtained
with the VLT mean that the measured equivalent widths were normalized by
their averaged value obtained in the frame of all spectral observations of
WD\,1953-011 exclusively with FORS1. The same was done for the results obtained
with the other telescopes. Because in all observing runs with the different
telescopes the observations were distributed more or less homogeneously over
the star's rotational cycle, such a normalization provided essential
unification of the data. Thus, hereineafter,
when mentioning $r_c$ and $EW$ we assume their normalized values.
This normalization makes it possible to consider all measurements
from the different spectral lines together. Results of the measurements are
presented in Table~\ref{tbl1}. Here we also present measurements from
spectral material obtained in the previous studies with Anglo-Australian
4-m telescope AAT \citep{MFMW00}, from the 8-m European telescope and
the 6-m Russian telescope \citep{VWB08}.

\subsection{Photometric measurements of WD\,1953-011}

The V-band CCD photometric observations were carried out in the
standard manner. For the comparison stars we used the same targets as used
by \citet{BM05}. Calibrating measured fluxes from WD\,1953-011
by fluxes from the standard stars we finally obtained a series
of $m_V$ values in stellar magnitudes at each of the individual
short-duration (a few minutes) exposures. The characteristic
uncertainty of the measurements is about 0.02 stellar magnitudes.

The rotation period of WD\,1953-011 ($P \approx 1.45$\,days)
and a preliminary inspection of the series of the $m_V$ values
allowed us to rule out a variability time-scale shorter than
a few thousand seconds. Hence we decided to consider a series of the magnitude
determinations where each individual value was obtained from the
weighted average of several consecutive points within equivalent exposure
bins of about 1500\,sec with two exceptions at
JD2454329.351 and JD2454329.442 \footnote{Due to weather conditions,
in these cases we had to average the data within about 5000\,sec
equivalent exposure bins, see Table~\ref{tbl2}}. Error bars have been
obtained as standard deviations from the averaged means. The measurements
are listed in Table~\ref{tbl2}.

\section{Results}

\subsection{Revising the rotation period}

The new observables ($m_V$, $r_c$, and $EW$) obtained for WD\,1953-011
from photometry and spectroscopy together with photometric data from
\citet{BM05} make it possible to revise the degenerate's rotation period
on a time base of about 10 years. All the observables used are 
variable with the star's rotation.

\begin{figure}
\vspace*{0.5cm}
\centering
\includegraphics[width=8.5cm, height=13.5cm, angle=0]{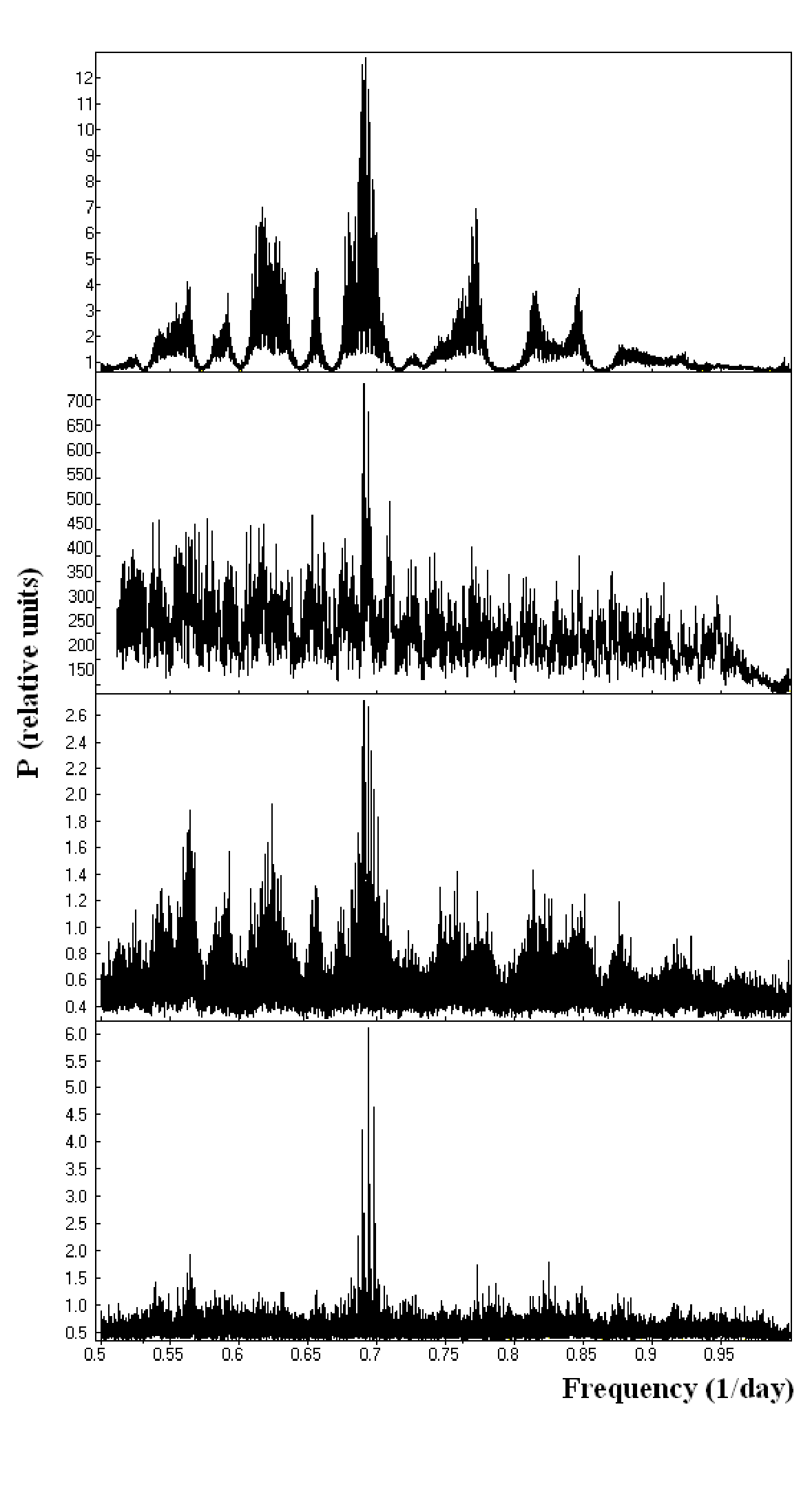}
\vspace*{0.1cm}
\caption{
Power spectra of the variation of (from top to bottom):
1. -- Stellar V-magnitude of WD\,1953-011 obtained from our
(discussed in this paper) observations. 2. -- Stellar V-magnitude of
WD\,1953-011 obtained from our observations together with those obtained by
\citet{BM05}. 3. -- Normalized H$\alpha$ and H$\beta$ equivalent
widths obtained from all available spectroscopic observations of WD\,1953-011
(i.e. from the observations discussed in this paper, from data
of \citet{MFMW00}, and from our previous observations presented by
\citet{VWB08}). 4. -- Normalized residual intensities of the H$\alpha$ and
H$\beta$ lines obtained from all available spectroscopic observations of
the degenerate.}
\label{fig1}
\end{figure}

To determine the rotation period we applied the Lafler-Kinman method
\citep{LK65} with user interface programmed by V.\,Goransky
(2004, private communication).
Power spectra of variation of the observables are presented in
Fig.~\ref{fig1}. The first plot (from top to bottom) presents
the periodogram of $m_V$ obtained from
our photometric observations. The second plot
is the periodogram obtained from our photometric observations together with
those obtained by \citet{BM05}. The third plot illustrates the power spectrum
of normalized H$\alpha$ and H$\beta$ equivalent widths. The
fourth plot is the power spectrum of the normalized residual intensity 
of the H$\alpha$ and H$\beta$ lines obtained from all available
spectroscopic observations of WD\,1953-011. As can be seen, all the spectra
exhibit a single strong peak at frequency of about 0.69\,day$^{-1}$
(P\,$\approx$\,1.45\,days).

Detailed study of the periodograms has shown that the most significant
and sinusoidal signal common to all observables corresponds to a period of
P~=~1.441788(6)~days. This period is very close to P~=~1.441769~days
found by \citet{BM05} from their photometry. Due to this agreement, here we
choose this period for phasing of all the data with the following
ephemeris (for which JD0 corresponds to the maximum of the light variation):

$$JD = 2,454,329.872 + 1.441788(6)\,E$$

\subsection{Phase-resolved photometry against Zeeman spectroscopy.}

In Fig.\,2 we illustrate the phase variations of the photometric magnitude
(the upper two panels on the figure), equivalent width of the
Balmer lines (the third panel from top) and residual intensity $r_c$
of the lines. All the data are phased with the ephemeris described above.
The lower two plots present the phase variation of the degenerate's global
surface magnetic field ($B_G$: mean field modulus integrated over the disc)
with the same ephemeris, and the projected fracional area $S$ of the
strong-field area on the disc (given in percent of the full disc area).
These data are taken from \citet{VWB08} and re-phased with the current
ephemeris. The two vertical dashed lines
in Fig.\,2 correspond to those phases ($\phi \approx 0.5;1.5$), when
the star exhibits the minimum brightness, and therefore maximum
projection of the dark area onto the disc.

As can be seen from Fig.\,2, the spectroscopic observable $r_c$ also
tends to have the smallest values at the phases of the light minimum.
According to the above explanations this suggests a direct relationship
between the dark area and strong-field spot.  However,
in comparison with the behaviour of the residual intensity, the variation
of equivalent width demonstrates a non-sinusoidal shape with
its absolute minimum shifted by $\Delta \phi \approx -0.05$ relative to the phase
of the light minimum ($\phi = 0.5$). Averaging $EW$ in the phase bins $0.41 < \phi < 0.47$
and $0.5 < \phi < 0.55$ and calculating the gradient of the data between these
two phases we have formal absolute minimum of the equivalent width change
to be at $\phi \approx 0.44 \pm 0.017$.
Besides, phased results of direct measurements of the projected
area $S$ (for details, see subsection 4.2 and Table\,3 in \citet{VWB08})
of the strong-magnetic spot on the disc seem to have the similar tendency,
shifted in phase by this value. Because of this we shall
discuss this phenonmenon in the present study in more detail.

This phase shift was first observed by \citet{Wad03} in their study of
WD\,1953-011 where they also estimated the ``visibility'' of the
strong-magnetic area via analysis of H$\alpha$ equivalent widths. This shift
could be produced by the presence of some morphologic difference in the
distribution of the field and photometric flux intensities within the
corresponding spots. Although the origin of the shift could be artificial
due to uncertainties in our simplifying assumption that the variation of
equivalent widths of the Balmer line profiles is attributed to
the strong-magnetic area only. The shift is, however, weak. And the general
behaviour of all the observables related to the dark and strong-field magnetic
areas exhibit correlated behaviour suggesting their geometric relationship.

Taking all of the above into account, we suppose that the dark and magnetic
spots are physically connected. In order to illustrate this conclusion,
below we model the dark spot from photometry comparing it
with the geometry of the strong-field area obtained by \citet{VWB08}.

\subsection{Modeling the dark/temperature spot}

In our model we assume that the star's flux variation is due to
surface inhomogeneity in the temperature distribution. Assuming a black body
relationship between temperature and energy flux we have:

\begin{equation}
m_V - \overline{m_V} = -2.512 \cdot lg \frac{F_V}{\overline{F_V}} =
-2.512 \cdot lg \frac{e^{  \frac {h c}{\lambda_v k \overline {T_{eff}}}} -1  }
{ e^{  \frac {h c}{\lambda_v k {T_{eff}}}} -1    }
\end{equation}

\noindent
where $h, c, k$ are the Plank constant, speed of light and Boltzman constant;
$\lambda_v$ is wavelength centered at the V-band filter;
$F_V$ and $\overline{F_V}$ are individual and averaged (over the entire
rotational cycle) photometric fluxes, while $T_{eff}$ and $\overline{T_{eff}}$ are individual
and averaged effective temperature respectively. Taking
$\overline{m_V} = 13\hbox{$.\!\!^{\rm m}$}632$ from our data, $\overline{T_{eff}} = 7920$\,K \citep{BLR01},

assuming that $e^{  \frac {h c}{\lambda_v k {T_{eff}}}} -
1 \approx e^{  \frac {h c}{\lambda_v k {T_{eff}}}} $ and weighting the
visible $T_{eff}$
by the line-of-sight flow of radiation as an average mean of the local surface
temperatures $T(\theta, \phi)$, we obtain the following approximation equation:

\begin{equation}
T_{eff} = \frac{\int\int T(\theta, \phi) \overline{n}
I(u,\overline{n}) sin \theta\, d \theta\, d \phi }{\int\int \overline{n}
I(u,\overline{n}) sin \theta\, d \theta\, d \phi }
= \frac{28926.765}{m_V - 9.98} = F(m_V)
\end{equation}

\noindent
where $\theta$ and $\phi$ are the spherical angles relative to the rotational
axis of the star; $\overline{n}$ is the cosine of the angle
between the surface element normal and the line of sight; and $u$ is the limb
darkening. For every
square element this cosine can be determined via the inclination $i$ of
the rotational axis with respect to the line-of-sight coordinates $\theta,\phi$
of the square and rotation phase $f$ as follows:

\begin{equation}
\overline{n} = cos\,i \cdot cos\,\theta + sin\,i \cdot sin\,\theta \cdot cos\,(\phi + f)
\end{equation}

\noindent
In the integration procedure we assume negative $\overline{n}$ to be zero
in order to integrate only over the visible hemisphere. Assuming a linear 
limb darkening law and dividing the
sphere by the elementary angles $\Delta \theta, \Delta \phi$,
equation (2) can be rewritten
as:

\begin{equation}
\Sigma\Sigma T_{l,n}\cdot A_{l,n,k} = F(m_V)_k
\end{equation}

\noindent
where $$A_{l,n,k} = \frac{\overline{n_{l,n,k}}
I_{l,n} sin \theta_l\,}{\Sigma\Sigma \overline{n_{l,n,k}} I_{l,n} sin \theta_l\, }
$$

\noindent
Here index $k$ corresponds to an observed rotation phase, and $l,n$ run
latitudinal and longitudinal positions of the elementary squares
in the star's rotational coordinate system.

Applying all
available photometric observations covering the full rotation cycle,
equation (2) can be solved by using any of non-linear least
square methods if the angle $i$ is known. In our model we used an
inclination of $i = 18^\circ$ as determined by \citet{VWB08}. With the
limb darkening $u = 0.45$, the temperature distribution up to 60
surface elements was calculated as presented below.

A general application of equation (4) to the analysis of nearly-sinusoidal signals
provides more than a single solution, including
physically unrealistic ones (negative surface temperatures for example).
In order to restrict the range of possible solutions for surface temperatures,
we modified the temperatures to have the following parametrized
form:

\begin{equation}
T(\theta, \phi) = \rm T_0 + F(x[\theta, \phi ])
\end{equation}

\noindent
where $F(x)$ is a positive function of the parameter $x$,
and ${\rm T_0 }$ is a physically reasonable lower temperature limit (say at
the center of the coolest area). To choose a reasonable $F(x)$
we tested several power, polynomial, and exponential functions. Finally,
the simplest case $F(x) = x^2$ was chosen. This case
provides satisfactory convergence of equation (4) if ${\rm T_0 }$ is fixed.

In such a formulation equation (4) was evaluated using the Marquardt
$\chi^2$ minimization method \citep{Bev69}. In our solution
we minimize the function:

\begin{equation}
\chi^2 = \Sigma (F(m_V)_k - \Sigma\Sigma T_{l,n}\cdot A_{l,n,k} ) ^ 2
\end{equation}

The used algorithm in FORTRAN-77 is presented by \citet{WBSW92}.

\subsection{Model results}

In Fig.\,3 we illustrate the tomographic portraits (left panels) of the
resulting smoothed temperature distribution patterns in comparison to the
strong-magnetic area. Dark areas correspond to low temperatures, and
brighter areas are higher temperatures. In these plots: the vertical axis is
latitude, the horizontal axis is longitude. The longitude is presented from
$-180^\circ$ to $+180^\circ$. Middle panels are corresponding phase
light curves (solid lines) obtained from the model fits of the photometric
observations (filled circles): vertical axis is $m_V$, horizontal axis is
rotation phase. Right panels illustrate in grey scale the
temperature distribution in the tomographic portraits. Five examples of
the star's temperature distribution with different T$_0$ are illustrated
(from top to bottom T$_0$~=~6500\,K;\,6750\,K;\,7000\,K;\,7100\,K;\,7120\,K).
The strong-field spot is schematically illustrated in Fig.\,3 by the elliptical
area bordered by a white solid line.

All the illustrated solutions demonstrate two
characteristic features: the presence of a temperature gradient
from the rotation pole to the equator, as well as the presence of a large low-temperature
spot (the shadow) covering 15-20\% of the star's hemisphere. The size of the shadow 
corresponds well to the size of the strong-field area estimated
by \citet{VWB08}. Three solutions at the lower temperature limits
$T_0 = 6500 K$, $T_0 = 6750 K$, and $T_0 = 7000 K$
(the upper three plots in Fig.\,3) present the most stable fits of the data
revealing temperature changes around the latitude of the strong-field
area location ($\theta \approx 67^\circ $). Higher temperature limits
shift the shadow to upper latitudes and provide less robust fits to the
photometric data (see the lower two plots in Fig.\,3). From consideration
of all the solutions we give the characteristic value to the lower temperature
limit (the tempereature at the center of the shadow) to be between
6000\,K and 7000\,K.

In order to find simpler solutions for the temperature distribution
we also examined $i$ as a parameter. The most stable and physically reasonable
solutions are found within the range $10^\circ \le i \le 20^\circ$.
Within this range the tomographic portraits demonstrate similar features
and temperature contrasts as discussed above.

The phase shift discussed previously ($\phi \approx 0.05 - 0.1$)
might be produced by a longitudinal shift between the dark and magnetic spots
of about $20^\circ$, and certainly not higher than $40^\circ$. However, even
in these extreme cases the dark and strong magnetic areas are
overlapped by more than 60\% . We therefore suggest that the dark and magnetic spots
are indeed physically connected.

The small difference in phasing of the corresponding observables
($m_V$ and $S$ in Fig.\,2) can also be explained by the following obvious
fact. In our previous investigation
of WD\,1953-011 \citep{VWB08} where we determined the rotationally modulated
projection $S$ of the strong field area to the line of sight we assumed a
constant temperature everywhere over the surface. Taking into account that
the center of the low-temperature region may have an effective temperature
about 20\% below the mean effective temperature of the photosphere,
we may suspect that these coolest parts of the surface
may by undetectable spectroscopically in the integrated hydrogen spectrum of
the degenerate. This could affect our analysis of the position of the
strong-field area and produce a small artificial shift in the position of the
strong-magnetic area.

The presence of the global gradient between the rotation pole to equator is also
a remarkable result. We find this gradient from practically all
tested T$_0$. The cases of T$_0 \approx 7100$\,K provide smaller values of the gradient.
Lower values of T$_0$ increase it. Also, testing smaller $i$
($11^\circ \le i \le 14^\circ$) we found some solutions with the smallest, but
nevertheless non-zero gradient. In all cases, the presence of the gradient
makes the shadow area have more diffusive edges, giving a smaller contrast in
temperatures between the shadow and neighboring areas. In this paper,
however, we are not making conclusions about the existence of the gradient,
but only discuss the implications.

\section{Discussion}

We have presented new photometric and spectral observations
of the magnetic white dwarf WD\,1953-011. From these
observations and those published previously we have re-determined
the star's rotation period, and studied the relationship between the brightness
variation and the variation of the strong-magnetic spot projection. We find a
new period of P~=~1.441788(6)~days. This period is consistent with all
the observables we employed, including those associated with the 
global magnetic field.

From direct comparison of the phased photometric and
spectroscopic observations we have found that the extremum of the photometric
variations of the V-band flux, and extremum of varying projection
of the magnetic spot to the line of sight, are shifted relative to each other
with phase shift of about 0.05 - 0.1. To study this shift we resolved
the phase brightness variation of the star assuming a thermal origin of
the photometric variability. We obtained a number of solutions which support
the idea that the variability comes from a low-temperature dark
spot covering about 15-20\% of the star's surface. The spot
might have a temperature characteristically lower than the mean by a factor of
about 20\%.

Comparing positions of the strong-magnetic area and dark spot we
conclude that they are located at the same latitudes, significantly
overlapping each other with a possible slight longitudinal shift that may be responsible for the phase shift discussed above.
Given that
the determination of the degenerate's strong-field area position and geometry
given by \citep{VWB08} were based on the assumption of constant effective
temperature in the atmosphere of WD\,1953-011, we propose that the observed shift may be a consequence of these assumptions. 
Besides, the total behaviour of all the
observables related to the dark and magnetic spots at other phases
are generally well-correlated within the error bars (see Fig.\,2). Due to
these arguments, we suggest that the association between the dark spot and
strong-field area in WD1953-011 is real.

The established association leads to the fundamental problem
related to origin and evolution of localized magnetic flux tubes in isolated
white dwarfs (see discussions in \citep{BM05,VWB08}). This idea, however,
requires additional observational evidence such as the presence of secular
drift of the tube which is still not established in WD\,1953-011. In this
context, we note that an independent search for the period using only the data related
to the global component of the star's magnetic field provides much better phasing
with a sightly longer period P~=~1.448~days \citep{VWB08}. This period
contradicts all the other observables related to strong-magnetic and
dark spots. Therefore this may admit the possibility that the
strong-magnetic area might slowly migrate over the star's surface.

The difference between the periods discussed above could be,
in principle, explained by the presence of a slow drift of the strong magnetic
area relative to the star's global magnetic field with a characteristic
time of about 280 days. This proposal, however, still requires additional
low-resolution spectropolarimetric, or high-resolution spectral observations
of the global field in WD\,1853-011.

Alternatively, in contrast to the idea of a low-temperature localized magnetic
flux tube, we might assume that the brightness/temperature distribution
pattern in WD\,1953-011 may have a correlation with the global field
intensity distribution, i.e., all local temperatures in WD\,1953-011 correlate
with corresponding local field intensities. In that case, the detected dark
spot is just the coolest part of the general temperature distribution
associated with the strongest magnetic features. This case might imply that
there is a common, ordered anisotropy of the light energy transfer in atmospheres
of magnetic degenerate stars due to the influence of their global
magnetic fields. Such a phenomenon would not be new for degenerate
stars. For example, it was discussed in application to neutron stars by
\cite{PS96}. For the magnetic white dwarfs, however, such a phenomenon has
not been observed until now.

\begin{acknowledgments}
Our thanks to L.~Ferrario, P.~Maxted, and C.~Brinkworth for providing
details of individual spectroscopic and photometric measurements of
WD\,1953-011. We are also especially grateful to our anonymous referee
for valuable comments and suggestions. GAW acknowledges Discovery Grant
support from the Natural Sciences and Engineering Research Council of Canada.
IH thanks KFICST (grant 07-179). S. Plachinda acknowledges support
from the Ukrainian Fundamental Research State Fund (M/364) and 
the Austrian Science Fund (P17890). LFM and MA acknowledge financial
support from the UNAM via PAPIIT grant IN114309. DS acknowledges financial
support from Deutsche Forschungsgemeinschaft (DFG), Research Grant
RE1664/7-1, and FWF Lise Meitner grant Nr. M998-N16.

\end{acknowledgments}



\begin{figure}
\vspace*{0.5cm}
\centering
\includegraphics[width=9.5cm, height=13.5cm, angle=0]{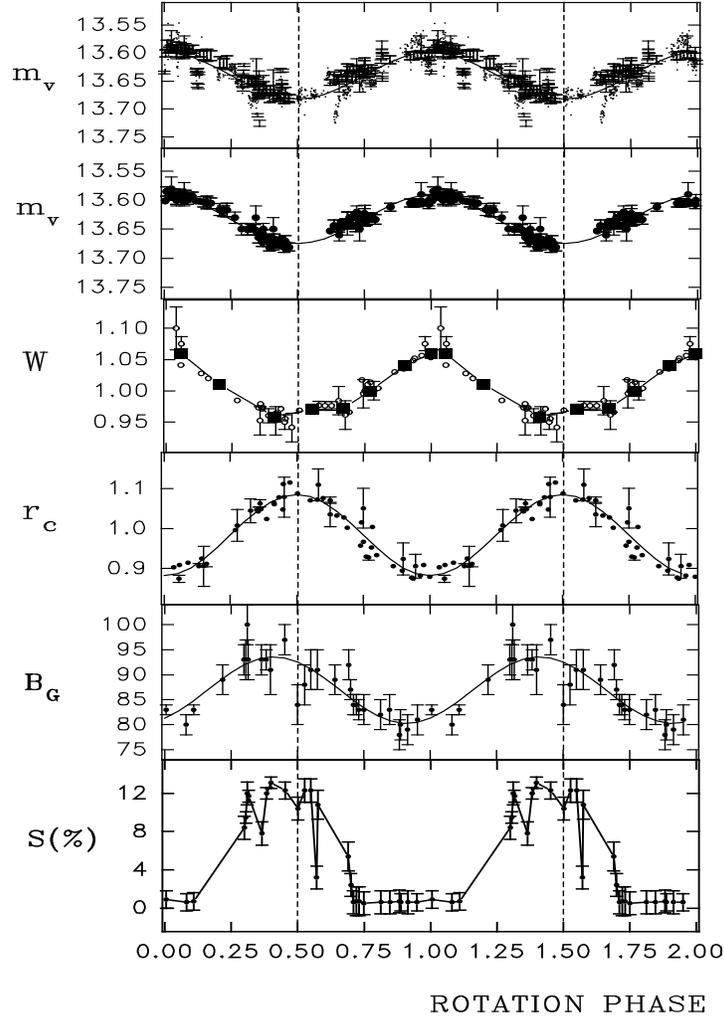}
\vspace*{0.1cm}
\caption{
Phase variations with rotation period P~=~1.441788(6)\,days of (from top to bottom):
1. -- Stellar V-magnitude of WD\,1953-011 obtained from our observations
together with those obtained by \citet{BM05}. 2. -- Stellar V-magnitude of
the degenerate obtained from our observations only.
3. -- Normalized H$\alpha$ and H$\beta$ equivalent
widths obtained from all available spectroscopic observations of WD\,1953-011
(i.e. from the observations discussed in this paper, from data
of \citet{MFMW00}, and from our previous observations presented by
\citet{VWB08}). 4. -- Normalized residual intensities of the H$\alpha$ and
H$\beta$ lines obtained from all available spectroscopic observations of
the degenerate. 5. -- Global magnetic field of the degenerate.
6.-- The projected fractional area of the strong-field area on the disc
(the data are taken from \citet{VWB08} and given in percents of full
disc square). The solid sinusoidal lines at plots 1,2,4,5 (from top to
bottom) are
least-square fits of the data. The solid squres at the middle plot 3 are
averaged data in bins. The two vertical solid lines pass through phases
0.5; 1.5. These phases correspond to minimum light energy from the star
and maximum projection of the magnetic spot to the line of sight.
}
\label{fig2}
\end{figure}

\begin{figure}
\vspace*{0.5cm}
\centering
\includegraphics[width=14.5cm, height=18.5cm, angle=0]{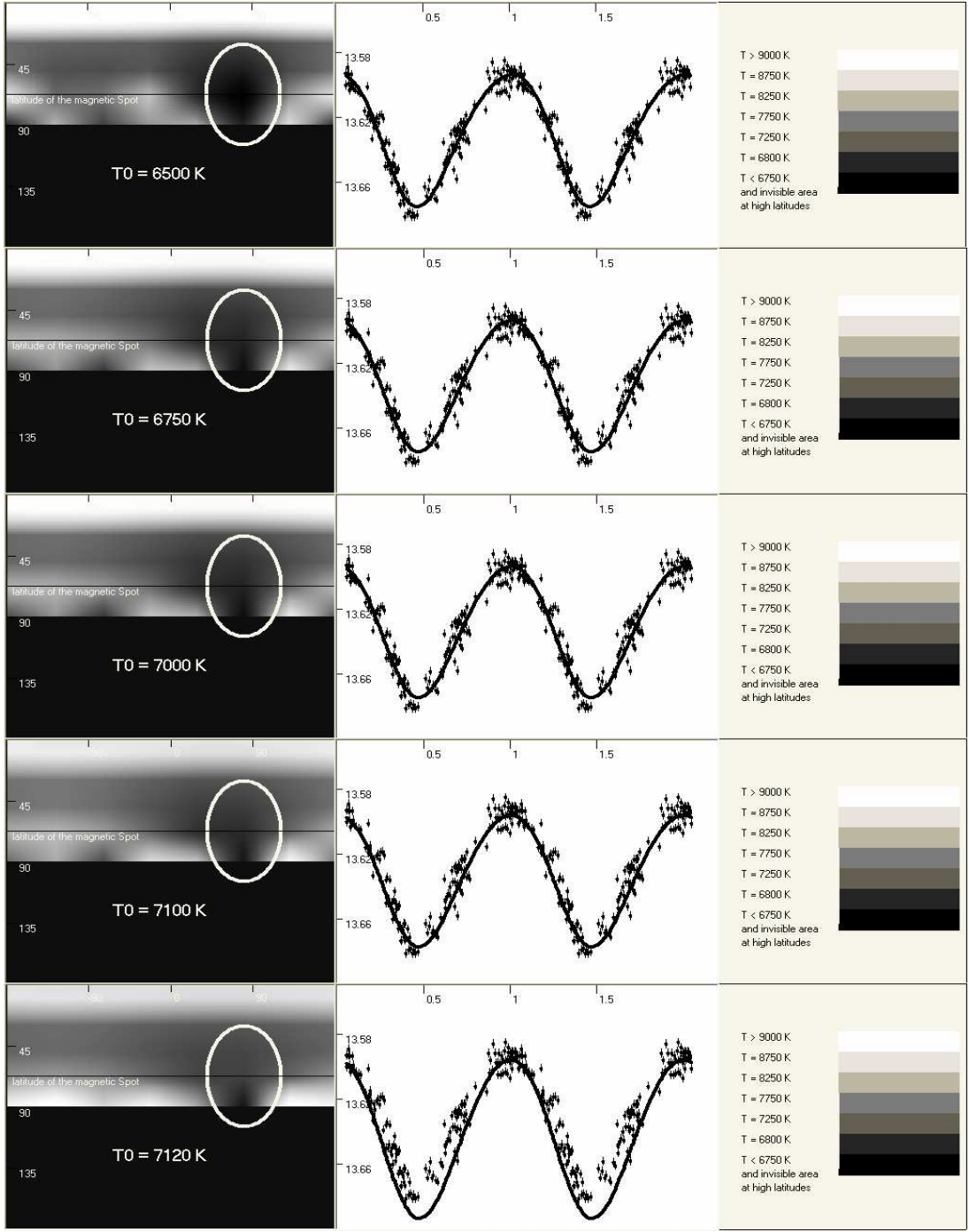}
\vspace*{0.1cm}
\caption{
Examples of the tomographic portraits (left panels) of the star's
temperature distribution in planar projections: vertical axis is latitude,
horizontal axis is longitude. The longitude is presented from
$-180^\circ$ to $+180^\circ$. Middle panels are corresponding phase
light curves (solid lines) obtained from model fits of the photometric
observations (filled circles): vertical axis is $m_V$, horisontal axis is
rotation phase. Right panels illustrate explanation to grey scale for the
temperature distribution in the tomographic portraits. Five examples of
the star's temperature distribution with different T$_0$ are illustrated
(from top to bottom T$_0$~=~6500\,K;\,6750\,K;\,7000\,K;\,7100\,K;\,7120\,K).}
\label{fig3}
\end{figure}

\newpage

\begin{table}
\caption{\label{tbl1} Spectral observations
of WD\,1953-011: column $JD$ is the Julian Date of the midpoint of the
observation; $Exp$ is the exposure time; $Telescope$ is the used telescope;
$r_c$ and $\sigma (r_c)$ are normalized residual intensity of spectral
line used in observations
(H$\beta$ in observations with ZTSH or H$\alpha$ in all other cases, see
explanations in the text) and associated error bar;
$EW$ and $\sigma(EW)$ are normalized equivalent width of the line
and its uncertainty correspondently.
}
\begin{tabular}{|c|c|c|c|c|c|c|}
\hline
\hline
$JD$  & $Exp$ (sec)&  $Telescope$ & $EW$ & $\sigma(EW) $ & $r_c$  & $\sigma(r_c)$ \\
\hline
2450676.955&   600&      AAT&     1.110&     0.007&      0.955&     0.004\\
2451391.948&   600&      AAT&     1.046&     0.007&      0.979&     0.004\\
2451391.955&   600&      AAT&     1.062&     0.009&      0.971&     0.004\\
2451391.962&   600&      AAT&     1.048&     0.006&      0.973&     0.004\\
2451392.059&  1800&      AAT&     1.077&     0.005&      0.971&     0.004\\
2451392.957&  1800&      AAT&     0.874&     0.009&      1.041&     0.004\\
2451393.066&  1800&      AAT&     0.905&     0.007&      1.027&     0.004\\
2451393.106&  1800&      AAT&     0.911&     0.007&      1.019&     0.004\\
2451393.943&  1200&      AAT&     0.957&     0.006&      1.017&     0.004\\
2451393.958&  1200&      AAT&     0.966&     0.005&      1.000&     0.004\\
2451393.973&  1200&      AAT&     0.929&     0.007&      1.010&     0.004\\
2451393.988&  1200&      AAT&     0.927&     0.007&      1.013&     0.004\\
2451394.003&  1200&      AAT&     0.951&     0.006&      0.994&     0.004\\
2452048.801&   840&      VLT&     0.874&     0.006&      1.051&     0.002\\
2452048.890&   840&      VLT&     0.879&     0.004&      1.053&     0.002\\
2452076.671&   840&      VLT&     0.996&     0.005&      0.984&     0.002\\
2452076.883&   840&      VLT&     1.060&     0.006&      0.964&     0.002\\
2452078.722&   840&      VLT&     1.001&     0.005&      0.965&     0.002\\
2452078.879&   840&      VLT&     0.933&     0.004&      1.004&     0.002\\
2452079.672&   840&      VLT&     1.046&     0.003&      0.973&     0.002\\
2452079.892&   840&      VLT&     1.086&     0.006&      0.968&     0.002\\
2452087.621&   840&      VLT&     0.905&     0.005&      1.030&     0.002\\
2452087.670&   840&      VLT&     0.894&     0.004&      1.035&     0.002\\
2452087.722&   840&      VLT&     0.878&     0.005&      1.048&     0.002\\
2452087.768&   840&      VLT&     0.881&     0.004&      1.056&     0.002\\
2452505.290&  3600&      BTA&     1.070&     0.007&      0.971&     0.007\\
2452505.327&  3600&      BTA&     1.072&     0.004&      0.976&     0.007\\
2452505.360&  3600&      BTA&     1.075&     0.005&      0.975&     0.007\\
2452505.397&  3600&      BTA&     1.070&     0.008&      0.975&     0.007\\
2454302.374&  8640&     ZTSH&     0.909 &    0.002 &     1.08 &    0.01 \\
2454302.489&  7477&     ZTSH&     0.909 &    0.002 &     1.08 &    0.01 \\
2454303.345&  7687&     ZTSH&     1.032 &    0.003 &     0.98 &    0.02 \\
2454304.405&  5656&     ZTSH&     1.024 &    0.003 &     1.01 &    0.01 \\
2454304.492&  5652&     ZTSH&     1.047 &    0.003 &     0.95 &    0.01 \\
2454316.358&  4082&     ZTSH&     1.028 &    0.004 &     0.96 &    0.02 \\
2454316.455&  2045&     ZTSH&     1.015 &    0.004 &     0.99 &    0.02 \\
\hline
\hline
\end{tabular}
\end{table}
\begin{table}
Table 1:--Continued:\\
\begin{tabular}{|c|c|c|c|c|c|c|}
\hline
\hline
$JD$  & $Exp$ (sec)&  $Telescope$ & $EW$ & $\sigma(EW) $ & $r_c$  & $\sigma(r_c)$ \\
\hline
2454316.513&  3897&     ZTSH&     1.004 &    0.004 &     1.04 &    0.03 \\
2454317.335&  5637&     ZTSH&     1.043 &    0.003 &     0.89 &    0.02 \\
2454317.421&  5612&     ZTSH&     1.065 &    0.003 &     0.89 &    0.02 \\
2454317.508&  5620&     ZTSH&     1.115 &    0.004 &     0.88 &    0.02 \\
2454318.319&  5593&     ZTSH&     0.903 &    0.003 &     1.05 &    0.03 \\
2454318.395&  3752&     ZTSH&     0.914 &    0.002 &     1.08 &    0.02 \\
2454318.471&  5581&     ZTSH&     0.925 &    0.003 &     1.09 &    0.02 \\
2454979.903&  5400&   2m SPM&     0.924 &    0.040 &                &               \\
2454979.969&  5400&   2m SPM&     0.905 &    0.030 &                &               \\
2454980.883&  5400&   2m SPM&     1.109 &    0.040 &                &               \\
2454980.949&  7200&   2m SPM&     1.035 &    0.030 &                &               \\
2454981.886&  5400&   2m SPM&     1.007 &    0.040 &                &               \\
2454981.957&  5400&   2m SPM&     1.044 &    0.030 &                &               \\
2454999.871&  9000&   2m SPM&     1.050 &    0.050 &                &               \\
2455000.885&  9000&   2m SPM&     1.078 &    0.050 &                &               \\
2455001.889&  9000&   2m SPM&     0.905 &    0.050 &                &               \\
\hline
\hline
\end{tabular}
\end{table}

\begin{table}
\caption{\label{tbl2} Photometric observations of WD\,1953-011:
column $JD$ is the Julian Date of the midpoint of the
observation, $Exp$ is the exposure time, columns m$_v$ and $\sigma(m_V)$
report stellar V-magnitude and its uncertainty respectively.
}
\begin{tabular}{c}
{\it A) Photometry with the 1.25\,m telescope AZT-11 at CRAO}\\
\end{tabular}
\begin{tabular}{|c|c|c|c||c|c|c|c|}
\hline
\hline
$JD$  & $Exp$ (sec)& m$_v$&$\sigma(m_V)$& $JD$  & $Exp$ (sec)& m$_v$&$\sigma(m_V)$\\
\hline
\hline
2454302.295& 1525& 13.605& 0.007&2454316.404& 1516& 13.628& 0.005\\
2454302.314& 1525& 13.605& 0.007&2454316.424& 1516& 13.631& 0.008\\
2454302.334& 1517& 13.604& 0.007&2454316.444& 1579& 13.627& 0.005\\
2454302.354& 1516& 13.603& 0.006&2454316.464& 1516& 13.623& 0.009\\
2454302.374& 1521& 13.604& 0.006&2454316.484& 1518& 13.630& 0.008\\
2454302.394& 1525& 13.607& 0.007&2454316.503& 1519& 13.629& 0.007\\
2454302.413& 1518& 13.601& 0.006&2454316.523& 1581& 13.633& 0.008\\
2454302.433& 1517& 13.594& 0.005&2454316.523& 1581& 13.633& 0.008\\
2454302.453& 1517& 13.594& 0.006&2454317.280& 1518& 13.650& 0.006\\
2454302.473& 1517& 13.600& 0.007&2454317.299& 1516& 13.650& 0.006\\
2454302.492& 1517& 13.589& 0.008&2454317.319& 1516& 13.652& 0.005\\
2454302.512& 1517& 13.602& 0.006&2454317.339& 1516& 13.660& 0.005\\
2454302.532& 1516& 13.599& 0.008&2454317.359& 1580& 13.658& 0.007\\
2454303.303& 1521& 13.653& 0.005&2454317.379& 1518& 13.666& 0.007\\
2454303.323& 1517& 13.644& 0.010&2454317.399& 1516& 13.664& 0.006\\
2454303.343& 1517& 13.645& 0.006&2454317.419& 1579& 13.667& 0.009\\
2454303.363& 1518& 13.637& 0.006&2454317.439& 1517& 13.673& 0.007\\
2454303.402& 1516& 13.638& 0.005&2454317.479& 1580& 13.671& 0.010\\
2454303.422& 1517& 13.642& 0.008&2454317.499& 1517& 13.681& 0.007\\
2454303.442& 1518& 13.644& 0.010&2454318.280& 1520& 13.585& 0.008\\
2454303.461& 1518& 13.633& 0.010&2454318.300& 1518& 13.591& 0.005\\
2454303.481& 1518& 13.634& 0.007&2454318.320& 1516& 13.591& 0.006\\
2454303.501& 1517& 13.635& 0.008&2454318.339& 1516& 13.591& 0.006\\
2454304.299& 1522& 13.650& 0.010&2454318.359& 1517& 13.591& 0.007\\
2454304.339& 1516& 13.651& 0.009&2454318.399& 1517& 13.593& 0.006\\
2454304.359& 1516& 13.664& 0.010&2454318.418& 1516& 13.593& 0.007\\
2454304.378& 1516& 13.673& 0.007&2454318.438& 1516& 13.601& 0.006\\
2454304.398& 1516& 13.671& 0.006&2454318.458& 1517& 13.602& 0.005\\
2454304.418& 1517& 13.681& 0.007&2454318.478& 1516& 13.603& 0.009\\
2454304.457& 1517& 13.678& 0.005&2454318.517& 1518& 13.605& 0.009\\
2454304.477& 1516& 13.679& 0.006&2454321.449& 1643& 13.616& 0.010\\
2454304.497& 1517& 13.681& 0.009&2454321.470& 1517& 13.619& 0.009\\
2454304.517& 1519& 13.680& 0.006&2454321.489& 1516& 13.617& 0.012\\
2454316.325& 1517& 13.648& 0.007&2454322.371& 1516& 13.612& 0.007\\
2454316.345& 1518& 13.646& 0.006&2454329.351& 5347& 13.635& 0.015\\
2454316.364& 1522& 13.640& 0.008&2454329.442& 5206& 13.627& 0.018\\
2454316.384& 1517& 13.630& 0.006&           &     &       &       \\
\hline
\hline
\end{tabular}
\end{table}
\begin{table}
\begin{tabular}{c}
Table 2--Continued:\\
{\it B) Photometry at OANSPM}
\end{tabular}

\begin{tabular}{|c|c|c|c||c|c|c|c|}
\hline
\hline
$JD$  & $Exp$ (sec)& m$_v$&$\sigma(m_V)$& $JD$  & $Exp$ (sec)& m$_v$&$\sigma(m_V)$\\
\hline
2454981.871& 700 &  13.63& 0.01&  2454996.895& 400 &  13.64& 0.02 \\
2454981.908& 700 &  13.65& 0.01&  2454996.947& 500 &  13.62& 0.01 \\
2454982.974& 700 &  13.59& 0.01&  2454997.845& 400 &  13.63& 0.02 \\
2454983.986& 600 &  13.65& 0.02&  2454997.885& 500 &  13.65& 0.01 \\
2454995.854& 500 &  13.59& 0.02&  2454997.938& 400 &  13.65& 0.02 \\
2454995.895& 400 &  13.60& 0.01&  2454998.857& 500 &  13.60& 0.01 \\
2454995.945& 400 &  13.58& 0.02&  2454998.898& 400 &  13.59& 0.02 \\
2454996.854& 500 &  13.66& 0.01&             &     &       &      \\
\hline
\hline
\end{tabular}
\end{table}

\end{document}